\documentclass[showpacs,preprintnumbers,amsmath,amssymb,prl,twocolumn,superscriptaddress]{revtex4}
\usepackage{graphicx}
\usepackage{bm}

\newcommand{\ve}[1]{\mathbf{#1}}

\newcommand{\im}{{\rm i}}
\newcommand{\xy}{3D$xy$\ }
\newcommand{\fig}[1]{Fig. \ref{#1}}
\def\comment#1{}

\begin{document}

\title{Critical properties of the $N$-color London model}

\author{J. Smiseth}
\affiliation{Department of Physics, Norwegian University of
Science and Technology, N-7491 Trondheim, Norway}
\author{E. Sm{\o}rgrav}
\affiliation{Department of Physics, Norwegian University of
Science and Technology, N-7491 Trondheim, Norway}
\author{A. Sudb{\o}}
\affiliation{Department of Physics, Norwegian University of
Science and Technology, N-7491 Trondheim, Norway}
\date{Received \today}

\begin{abstract}
The critical properties of $N$-color London model  
are studied in $d=2+1$ dimensions. The model is dualized
to a theory of $N$ vortex fields interacting through a Coulomb and a screened 
potential. The model with $N=2$ shows two anomalies in the specific heat. From the 
critical exponents $\alpha$ and $\nu$, the mass of the gauge field, and the vortex 
correlation functions, we conclude that one anomaly corresponds to an
{\it inverted} \xy  fixed point, while the other corresponds to a  \xy 
fixed point. There are  $N$ fixed points, namely one 
corresponding to an inverted \xy fixed point, and $N-1$ corresponding to neutral \xy 
fixed points. This represents a novel type of quantum fluid, where superfluid
modes arise out of charged condensates.
\end{abstract}


\maketitle
Ginzburg-Landau (GL) theories with several complex scalar matter fields minimally coupled
to one gauge field are of interest in a wide variety of systems, such as multiple 
component (color) superconductors, metallic phases of light atoms such 
as hydrogen \cite{egor2002,Ashcroft1999}, and as effective theories for easy-plane quantum 
antiferromagnets  \cite{senthil2003,motrunich2004,sachdev2004}. 
The model also is highly relevant in particle physics 
where it is called two-Higgs doublet model \cite{tdlee}.
In metallic hydrogen
the scalar fields represent Cooper pairs of electrons and protons,
which excludes the possibility of inter-color pair tunneling, i.e.
 there is no Josephson coupling between  different components of the condensate. 
The same two-color action in $(2+1)$ dimensions, where the matter
fields originate in a bosonic representation of spin operators, is claimed 
to be the critical sector of a field theory separating a N{\'e}el state 
and a paramagnetic (valence bond ordered) state of a two dimensional 
quantum antiferromagnet at zero temperature with an easy-plane
anisotropy present \cite{senthil2003,sachdev2004}. This happens because, 
although the effective description of the antiferromagnet involves an  
{\it a priori} compact gauge field, it must be supplemented by Berry-phase 
terms in order to properly describe $S=1/2$ spin systems 
\cite{haldane1988,read_sachdev}. Berry-phases cancel the 
effects of monopoles at the critical point \cite{senthil2003,sachdev2004}. 
In this paper, we point out novel physics of the quantum fluid that
arises out of an $N$-color charged condensate when no intercolor
Josephson coupling is present.

For a  detailed analysis of the phase transitions in such a generalized 
GL model, we study an $N$ component GL theory in $(2+1)$ dimensions 
with no Josephson coupling term. The model is defined by $N$ complex 
scalar fields $\{\Psi^{(\alpha)}(\ve r) \ | \ \alpha=1\dots N\}$ coupled 
through the charge $e$ to a fluctuating gauge field $\ve{A}(\ve r)$, with  
Hamiltonian 
\begin{equation}
\label{gl_action}
H = \sum_{\alpha=1}^N \frac{|(\nabla -\im e\ve{A})\Psi^{(\alpha)}|^2}{2 M^{(\alpha)}} 
+ V(\{\Psi^{(\alpha)}\}) + \frac{1}{2}(\nabla\times\ve{A})^2
\end{equation}
where $M^{(\alpha)}$ is the $\alpha$-component condensate mass.
The potential $V(\{\Psi^{(\alpha)}(\ve r)\})$ is assumed to be only a
function of $|\Psi^{(\alpha)}(\ve r)|^2$. 
The model is studied in the phase-only (London) approximation $\Psi^{(\alpha)}(\ve
r) = |\Psi^{(\alpha)}_0|\exp [ \im\theta^{(\alpha)}(\ve r) ]$ and is 
discretized on a lattice with spacing $a=1$ \cite{Dasgupta}. In the 
Villain approximation the partition function reads 
\begin{equation}
\begin{split}
  \label{villain}
Z =& \int_{-\infty}^{\infty}\mathcal{D}\ve A\prod_{\gamma=1}^N\int_{-\pi}^{\pi}\mathcal{D}\theta^{(\gamma)}\prod_{\eta=1}^N\sum_{\ \ve n^{(\eta)}}\exp(-S) \\
S =& \sum_{\ve r}\left(\sum_{\alpha=1}^N\frac{\beta|\Psi^{(\alpha)}_0|^2}{2 M^{(\alpha)}}(\Delta\theta^{(\alpha)} - e\ve{A} +2\pi \ve n^{(\alpha)})^2\right. \\
&\left.+\frac{\beta}{2}(\Delta\times\ve{A})^2\right),
\end{split}
\end{equation}
where  $\ve n^{(\alpha)}(\ve r)$ are integer vector fields
ensuring $2\pi$ periodicity, and the lattice position index vector $\ve r$ of the
fields is suppressed. The symbol $\Delta$ denotes the lattice difference operator 
and $\beta=1/T$ is the inverse temperature. {\it Here, we stress the importance of keeping 
track of the $2\pi$ periodicity of the individual phases}.  The kinetic 
energy terms are linearized by introducing $N$ auxiliary fields $\ve
v^{(\alpha)}$. Integration over all $\theta^{(\alpha)}$ produces the
local constraints $\Delta\cdot\ve v^{(\alpha)}=0$, which are
fulfilled by the replacement $\ve v^{(\alpha)} \to \Delta\times\ve
h^{(\alpha)}$. We recognize $\ve h^{(\alpha)}$ as the dual gauge
fields of the theory. By fixing the gauge $n^{(\alpha)}_z=0$ and
performing a partial integration we may introduce 
the vortex fields $\ve m^{(\alpha)} = \Delta\times\ve n^{(\alpha)}$.
We integrate out the gauge field $\ve A$ and get a theory in the dual gauge fields 
$\ve h^{(\alpha)}$ and the vortex fields $\ve m^{(\alpha)}$ where $\Delta\cdot\ve m^{(\alpha)}=0$
\begin{equation}
\begin{split}  \label{dual2}
S = \sum_{\ve r}&\left[2\pi\im\sum_{\alpha=1}^N \ve m^{(\alpha)}\cdot\ve h^{(\alpha)}+ 
\sum_{\alpha=1}^N\frac{(\Delta\times\ve h^{(\alpha)})^2}{2\beta|\psi^{(\alpha)}|^2}\right.\\
+&\left. \frac{e^2}{2\beta}\left(\sum_{\alpha=1}^N\ve h^{(\alpha)}\right)^2\right],
\end{split}
\end{equation}
where $|\psi^{(\alpha)}|^2 = |\Psi^{(\alpha)}_0|^2/M^{(\alpha)}$.
Note how the {\it algebraic sum} of the dual photon
fields is massive. This differs from 
the case $N=1$, where $e$ produces one massive dual photon with bare
mass $e^2/2$, and the model describes a vortex field $\ve m$ interacting 
through a \textit{massive} dual gauge field $\ve h$. However, when $N\ge 2$, 
since $\Delta\cdot\ve
m^{(\alpha)}=0$, a gauge transformation $\ve h^{(\alpha)}\to \ve h^{(\alpha)}
+ \Delta g^{(\alpha)}$ for $\alpha=1\dots N$ leaves the action
invariant if one of the gauge fields, say $\ve h^{(\eta)}$ compensates
the sum in the last term in (\ref{dual2}) with $\Delta g^{(\eta)} =
-\sum_{\alpha\neq\eta}\Delta  g^{(\alpha)}$. 

Integrating out the dual gauge fields we get a generalized theory of $N$ interacting
vortex fields
\begin{equation}
  \label{vortex_action}
\begin{split}
Z &= \sum_{\ve m^{(1)}}\cdots\sum_{\ve m^{(N)}}\delta_{\Delta\cdot\ve m^{(1)},0}\cdots\delta_{\Delta\cdot\ve m^{(N)},0}
\times e^{- S_V }\\
S_V & =\sum_{\ve r,\ve r^\prime}\sum_{\alpha,\eta}\ve m^{(\alpha)}(\ve r)
D^{(\alpha,\eta)}(\ve r - \ve r^\prime)\ve m^{(\eta)}(\ve r^\prime)
\end{split}
\end{equation}
where $\delta_{x,y}$ is the Kronecker-delta, and the vortex
interaction potential $D^{(\alpha,\eta)}(\ve r)$ is the inverse
discrete Fourier transform of $\widetilde{D}^{(\alpha,\eta)}(\ve q)$, where
\begin{equation}
\label{potential}
\frac{\widetilde{D}^{(\alpha,\eta)}(\ve q)}{2\pi^2\beta|\psi^{(\alpha)}|^2} = 
\frac{\lambda^{(\eta)}}{|\ve Q_{\ve q}|^2 + m_0^2} + \frac{\delta_{\alpha,\eta} 
- \lambda^{(\eta)}}{|\ve Q_{\ve q}|^2},
\end{equation}
$\lambda^{(\alpha)} = |\psi^{(\alpha)}|^2/\psi^2$ and $\psi^2 = \sum_{\alpha=1}^N|\psi^{(\alpha)}|^2$. 
Here, $m^2_0 = e^2 \psi^2$ is the square of the bare inverse screening length
in the intervortex interaction, and $|\ve Q_{\ve q}|^2$ 
is the Fourier representation of the lattice Laplace operator. The first term of the vortex 
interaction potential (\ref{potential}) is a Yukava screened potential, {\it while the second term 
mediates long range Coulomb interaction between 
vortex fields}. If $N=1$ the latter cancels out exactly and we are left with the well studied 
vortex theory of the GL model which has a charged fixed point for $e \neq 0$ 
\cite{Herbut_Tesanovic1996,hove2000}. For 
$N\ge2$ we find a theory of vortex loops of  $N$ colors interacting through long range 
Coulomb interaction. If $N$ grows 
to infinity, $\psi^2\to\infty$ and the vortex fields
interact via a diagonal unscreened $N \times N$ Coulomb matrix. This reflects 
the inability of one single gauge field $\bf{A}$ to screen a large
number of vortex species. {\it The case $N \geq 2$ has features
with no counterpart in the case $N=1$ \cite{Dasgupta,hove2000}, namely
neutral superfluid modes arising out of charged condensates}.

The above vortex system may be formulated as a field theory, introducing $N$ complex 
matter fields $\phi^{(\alpha)}$ for each vortex species, minimally coupled to the dual 
gauge fields $\bf{h}^{(\alpha)}$. This generalizes the dual theory
for $N=1$ pioneered in \cite{kleinert_book}. The theory reads
(see also \cite{sachdev2004})
\begin{equation}
\label{dual_action}
\begin{split}
S_{\rm{dual}}  =& \sum_{\ve r} \left[ \sum_{\alpha=1}^N \left( m_\alpha^2 |\phi^{(\alpha)}|^2 + 
|(\Delta - \im \ve h^{(\alpha)})\phi^{(\alpha)}|^2 \right.\right.\\
&+ \left.\frac{(\Delta\times\ve h^{(\alpha)})^2}{2\beta|\psi^{(\alpha)}|^2} \right) 
 +  \frac{e^2}{2\beta} \left( \sum_{\alpha=1}^N\ve h^{(\alpha)}
 \right)^2 \\
&+ \left.\sum_{\alpha,\eta} g^{(\alpha, \eta)} |\phi^{(\alpha)}|^2 |\phi^{(\eta)}|^2 \right].
\end{split}
\end{equation}
Here, we have added chemical potential (core-energy) terms for the vortices, as well 
as steric  short-range repulsion interactions between vortex
elements. In the $N=1$ case, a RG treatment of the mass term of the dual gauge 
field yields $\partial e^2/\partial \ln l = e^2$, and hence this term
scales up, suppressing the dual gauge field. 
Correspondingly, for $N \geq 2$, this suppresses $\sum_{\alpha} \ve h^{(\alpha)}$, but
not each individual dual gauge field. For the particular case $N=2$,
assuming the same to hold, we end 
up with a gauge theory of two complex matter fields coupled minimally to one massless 
gauge field, which was also precisely the starting point. Thus
the theory is self-dual for $N=2$ \cite{motrunich2004,sachdev2004}. 
For $N=1$, it is known that a charged theory in $d=2+1$ dualizes into a 
$|\phi|^4$ theory and vice versa \cite{hove2000}.
The vortex tangle of the \xy model is incompressible and the dual theory 
is a massless gauge theory such that $\langle \phi \rangle \neq 0$  
is prohibited. For $e² \neq 0$, the dual theory has global 
symmetry,  and vortex condensation and 
 $\langle \phi \rangle \neq 0$ is possible \cite{hove2000}. 

For  $N=2$, Monte Carlo (MC) simulations have been carried out for the action (\ref{vortex_action}) with 
parameters $|\psi^{(1)}|^2 = 1/2$, $|\psi^{(2)}|^2 = 1$, $e^2=1/4$, and $m_0^2 = 3/8$. 
Here,  $|\psi^{(1)}|^2$ and $|\psi^{(2)}|^2$ have been chosen to have well-separated bare 
energy scales associated with the twist of the two types of phases, and $m_0$
has been chosen to be of the order of the inverse lattice spacing in the problem to
avoid difficult finite-size effects. One MC update consists of inserting  
elementary vortex loops of random direction and species according to the Metropolis algorithm.

\begin{figure}[htb]
\centerline{\scalebox{0.34}{\rotatebox{-90.0}{\includegraphics{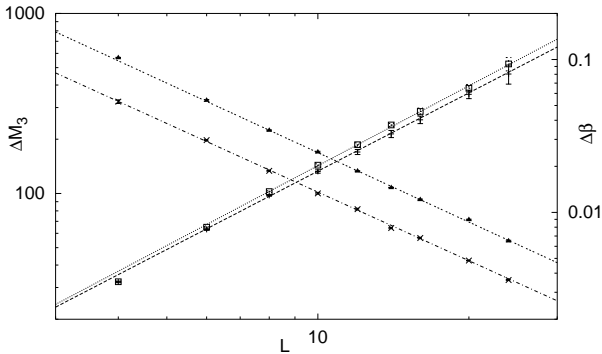}}}} 
\caption{\label{m3} The FSS of the peak to peak value of
  the third moment $\Delta M_3$ labeled ($\Box$) and (+) for
  $T_{\rm c1}$ and $T_{\rm c2}$ respectively. The scaling of
  the width between the peaks $\Delta\beta$ labeled ($\blacktriangle$) and ($\times$) for
  $T_{\rm c1}$ and $T_{\rm c2}$ respectively. The lines
are power law fits to the data for $L>6$ used to extract $\alpha$ and $\nu$.}
\end{figure}

We observe two anomalies in the specific heat at 
$T_{\rm c1}$ and $T_{\rm c2}$ where $T_{\rm c1}< T_{\rm c2}$ . 
We find $T_{\rm c1}$ and $T_{\rm c2}$ from scaling of the second moment of 
the action  $\langle (S_V - \langle S_V \rangle )^2 \rangle$ to be 
$T_{\rm c1} = 1.4(6)$ and $T_{\rm c2} = 2.7(8)$. 
To check the 
criticality of these anomalies we have
calculated the critical exponents $\alpha$ and $\nu$ by applying finite
size scaling (FSS) of 
 $M_3=\langle(S_V-\langle S_V \rangle)^3\rangle$ \cite{m3}. The 
peak to peak value of this quantity scales with system size $L$ as $L^{(1+\alpha)/\nu}$, 
the width between the peaks scales as $L^{-1/\nu}$. The advantage of this
 is that asymptotically correct
behavior is reached for practical system sizes. The FSS plots for system sizes 
$L=4, 6, 8, 10, 12, 14, 16, 20, 24$ are shown in \fig{m3}.
From the scaling we conclude that both anomalies are in fact critical points, and 
we obtain $\alpha = -0.02 \pm 0.02$ and $\nu=0.67\pm
0.01$ for $T_{\rm c1}$ and $\alpha = -0.03 \pm 0.02$ and $\nu = 0.67\pm 0.01$ for
	 $T_{\rm c2}$. These values are consistent with those of the
\xy and the {\it inverted} \xy universality classes found with high
precision to be $\alpha = -0.0146(8)$ and $\nu =0.67155(3)$
\cite{kleinert1999}.

To characterize these phase transitions further, we consider 
$\mathcal{G}_{\ve A}(q) = \langle \ve A_q\cdot\ve
A_{-q}\rangle$ and  $\mathcal{G}_{\Sigma\ve
  h}(q) = \langle (\sum_{\alpha}\ve h^{(\alpha)}_q)\cdot(\sum_{\alpha}\ve h^{(\alpha)}_{-q})\rangle$, 
expressed in terms of  
$G^{(+)}(q) = 
\langle |\sum_{\alpha}|\psi^{(\alpha)}|^2\ve m^{(\alpha)}_q|^2 \rangle$
as
\begin{equation}
  \label{gaugeprop}
\begin{split}
 &\mathcal{G}_{\ve A}(q)  = \frac{2/\beta}{|\ve Q_{\ve q}|^2 + m_0^2}\left(1 +\frac{2\pi^2\beta
  m_0^2}{|\ve Q_{\ve q}|^2 }\frac{G^{(+)}(q)}{|\ve Q_{\ve  q}|^2 + m_0^2}\right) \\
 &\mathcal{G}_{\Sigma\ve h}(q) = 
\frac{2\beta\psi^2}{|\ve Q_{\ve q}|^2 + m_0^2}\left(1  -  \frac{2\pi^2\beta}{\psi^2}\frac{G^{(+)}(q)}{|\ve Q_{\ve q}|^2 + m_0^2}
  \right).
\end{split}
\end{equation}
The masses of  $\ve A$ and $\sum_{\alpha} \ve h^{(\alpha)}$ are defined by 
$m^2_{\ve A}=\lim_{q\to0}2\mathcal{G}_{\ve  A}(q)^{-1}/\beta$ and 
$m^2_{\Sigma\ve h}=\lim_{q\to0}2\beta\psi^2 \mathcal{G}_{\Sigma\ve h}(q)^{-1}$.

We briefly review the case $N=1$ \cite{hove2000}. The dual field 
theory of the neutral fixed point ($m_0^2=0$) is a charged theory describing an
incompressible vortex tangle. The leading behavior of the vortex 
correlator is $\lim_{q\to 0} 2 \pi^2\beta G^{(+)}(q)\sim[1-C_2(T)]q^2$,
$q^2-C_3(T)q^{2+\eta_{\ve h}}$, and $q^2+ C_4(T)q^4$ for $T<T_{c}$,
$T=T_{\rm c}$, and $T>T_{\rm c}$ respectively. For $T < T_{\rm c}$
we have $m^2_{\Sigma\ve h} =0$ ($N=1$),
however for $T>T_{\rm c}$ the $1/q^{2}$  terms in $\mathcal{G}_{\Sigma\ve h}(q) $ cancel out exactly and
this mass attains an expectation value. 
At the charged fixed point $(m_0^2 \neq 0$) of the GL model, the effective field 
theory of the vortices is a neutral theory. The vortex tangle is compressible
with a scaling ansatz for the vortex correlator $\lim_{q\to 0}G^{(+)}(q)\sim
q^2$, $q^{2-\eta_{\ve A}}$, and $c(T)$ for $T<T_{\rm c}$, 
$T=T_{\rm  c}$, and
$T\geq T_{\rm c}$, respectively. Consequently, from
(\ref{gaugeprop}), the mass $m_{\ve A}$ drops to zero at $T_{\rm c}$, and 
the mass of the dual gauge field $m_{\ve h}$ is
finite for all temperatures and has a kink at $T_{\rm c}$. Renormalization group 
arguments yield $\eta_{\ve  A}=4-d$ where $d$ is the dimensionality 
\cite{HLM1974,Herbut_Tesanovic1996}, which has recently been verified numerically 
\cite{hove2000,kajantie2004}.

\begin{figure}[htb]
\centerline{\scalebox{0.34}{\rotatebox{-90.0}{\includegraphics{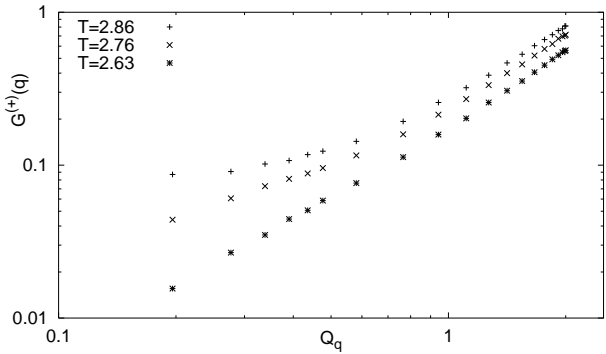}}}} 
\caption{\label{G_q} $G^{(+)}(q)$ for $N=2, L=32$. For 
$T=2.86 > T_{\rm{c2}}$, $T=2.76\simeq T_{\rm c2}$, and $T=2.63 < T_{\rm{c2}}$,  
$\lim_{\bf{q} \to 0} G^{(+)}(q) \sim c(T)$,  $\sim q$, and  $\sim q^2$, 
respectively.}
\end{figure}
The vortex correlator for  $N=2$  is sampled in real space and  $G^{(+)}(q)$ is 
found by discrete Fourier transformation, it is shown in Fig. (\ref{G_q}).  At 
$T=T_{\rm c1}$ the leading behavior is $G^{(+)}(q)\sim q^2$ on both sides of 
$T_{\rm c1}$. Consequently, due to (\ref{gaugeprop}), $m_{\ve A}$ and
$m_{\Sigma \ve h}$ are finite in this regime. This shows that the
vortex tangle is incompressible and that the anomalous scaling
dimension $\eta_{\ve A}=0$, which corresponds to a neutral fixed
point. 
Below $T_{\rm c2}$ the dominant behavior is $G^{(+)}(q)\sim q^2$ whereas 
$G^{(+)}(q)\sim c(T)$ above $T_{\rm c2}$. At $T=T_{c2}$, $G^{(+)}(q)\sim q$ 
indicating $\eta_{\ve A}=1$. Accordingly, $m_{\ve A}$ is finite below 
$T_{\rm c2}$ and zero for $T \ge T_{\rm c2}$.

\begin{figure}[htb]
\centerline{\scalebox{0.34}{\rotatebox{-90.0}{\includegraphics{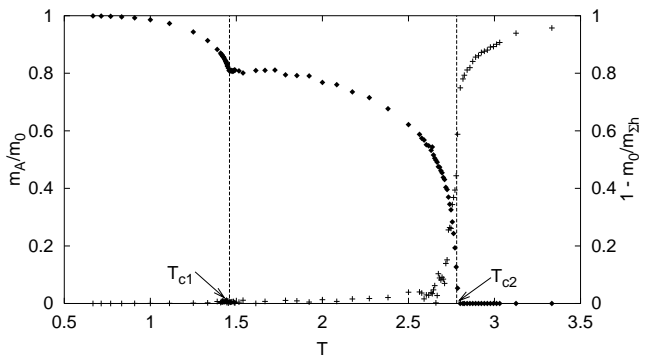}}}} 
\caption{\label{gauge_mass} The mass $m_{\ve A}$ ($\blacklozenge$) and
  $1-m_0/m_{\Sigma\ve h}$ (+) found from Eq. (\ref{gaugeprop}). Two  non-analyticities 
  can be seen in $m_{\ve A}$ at $T_{\rm c1}$ and $T_{\rm c2}$, corresponding a neutral 
  fixed point and a charged Higgs fixed point, respectively. An abrupt increase in 
  $m_{\Sigma\ve h}$ due to vortex condensation is located at $T_{\rm c2}$. } 
\end{figure}

For  $T \alt T_{\rm c2}$, $m_{\ve A}$  scales according to
$\mathcal{G}_{\ve A}(q)^{-1}\frac{2}{\beta} = m_{\ve A}^2 +
Cq^{2-\eta_{\ve A}} + \mathcal{O}(q^\delta)$ for small $q$ where $\delta > 2-\eta_{\ve A}$ \cite{kajantie2004},  
with a corresponding Ansatz for $\mathcal{G}_{\Sigma\ve h}(q) $. For each
coupling we fit $\mathcal{G}_{\ve A}(q)^{-1}$ data from system sizes
$L=8,12,20,32$ to the Ansatz. The results for $m_{\ve A}$ (and
$m_{\Sigma \ve h}$, found similarly), are given in
\fig{gauge_mass}. The system exhibits Higgs mechanism
at $T=T_{\rm c2}$ when $m_{\ve A}$ drops to zero. Furthermore $m_{\ve A}$
has a kink at $T_{\rm c1}$ due to ordering of $\theta^{(1)}$.
The anomalies 
in $m_{\ve A}$ coincide precisely with $T_{\rm c1}$ and $T_{\rm c2}$
determined from scaling of $\langle (S_V - \langle S_V \rangle)^2 \rangle$. 
Note also how $m_{\Sigma\ve h}$ changes abruptly at $T_{\rm c2}$.
This is due to a sudden change in screening of 
$\sum_{\alpha=1}^N\ve h^{(\alpha)}$ by the vortex-loop proliferation at $T=T_{c2}$, 
giving an abrupt increase in 
$m_{\Sigma\ve h}$, analogously to what happens for $N=1, e \neq 0$ \cite{hove2000}. 
Above  $T_{\rm{c2}}$,  $\bf{A}$ is massless, giving a compressible vortex 
tangle which accesses configurational entropy better than an 
incompressible one. Below $T_{\rm{c2}}$, ${\bf{A}}$ is massive and 
merely renormalizes $|\Psi|^4$ terms in Eq. (\ref{gl_action}). The theory 
is effectively a $|\Psi |^4$ theory in this regime. Thus, the remaining 
proliferated vortex species originating in the matter fields with lower 
bare stiffnesses form  vortex tangles  as if they originated in a neutral 
superfluid. For the general $N$ case, a Higgs mass is generated at the highest 
critical temperature, after which  ${\bf{A}}$ merely renormalizes the $|\Psi|^4$ 
term, such that the Higgs  fixed point is followed by $N-1$ neutral fixed points 
as the temperature is lowered.

We now discuss the vortex mode $\ve m^{(1)}  - \ve m^{(2)}$, demonstrating 
that it should be identified as a superfluid mode in the system.
Its properties are controlled by $\mathcal{G}_{\Delta \ve h} (q) \equiv
\langle |\ve h^{(1)}_{q}-\ve h^{(2)}_{q}|^2 \rangle$.
A dual Higgs phenomenon for $N=2, T=T_{c1}$  involving  
$\mathcal{G}_{\Delta \ve h} (q) $
may be demonstrated as follows.  
Introducing $G^{(-)}(q) =  \langle |\ve m^{(1)}_q  - \ve m^{(2)}_q|^2 \rangle$ and $G^{(\rm
  m)}(q) = \langle (\ve m^{(1)}_q  - \ve m^{(2)}_q) \cdot
(\sum_{\alpha=1}^2|\psi^{(\alpha)}|^2\ve m^{(\alpha)}_{-q}) \rangle$
we find, in the notation used in Eqs. (\ref{potential}) and (\ref{gaugeprop})
\begin{widetext}
\begin{equation}
\mathcal{G}_{\Delta\ve h}(q) =
\frac{8 \beta \lambda^{(1)}\lambda^{(2)}\psi^2}{|\ve Q_{\ve q}|^2 }
 \left\{ 1 - \frac{2\pi^2\beta\lambda^{(1)}\lambda^{(2)}\psi^2 G^{(-)}(q)}{|\ve Q_{\ve q}|^2}
- \frac{2\pi^2\beta(\lambda^{(1)} - \lambda^{(2)}) G^{(m)}(q)}{|\ve Q_{\ve q}|^2 + m_0^2} 
\right\}  + (\lambda^{(1)} - \lambda^{(2)})^2\mathcal{G}_{\Sigma\ve h}(q).
\label{hm_dual}
\end{equation}
\end{widetext}
The $G^{(-)}(q)$ correlation function is always $\sim  q^2,  q \to 0$, but 
has a nonanalytic coefficient of $ q^2$, determined by the helicity 
modulus $\Upsilon$ of the neutral mode $\ve m^{(1)} - \ve m^{(2)}$. When $\Upsilon$
vanishes at $T_{c1}$ through a disordering of $\theta^{(1)}$, thus destroying
the superfluid neutral mode, the first and second term in the bracket cancel, which in turn 
cancels the $1/ q^2$ term in $ \mathcal{G}_{\Delta\ve h}(q)$. This 
produces a dual Higgs mass $m_{\Delta \ve h}$ defined by  
$\mathcal{G}_{\Delta\ve h}(q) \sim 1/(q^2 + m^2_{\Delta \ve h})$ for $T > T_{c1}$. The remaining terms in Eq. 
(\ref{hm_dual}) contribute to determining the actual value of $m_{\Delta \ve h}$. Thus, while 
$\ve h^{(1)} + \ve h^{(2)}$ is always massive, cf. Eq. (\ref{dual2}), $\ve h^{(1)} - \ve h^{(2)}$ 
is massless below $T_{\rm c1}$ and massive above $T_{\rm c1}$. Therefore $\ve h^{(1)} - \ve h^{(2)}$ 
plays the role of a gauge degree of freedom, providing a dual counterpart to $\ve A$ in 
Eq. (\ref{gl_action}). This is evident when $|\psi^{(1)}|^2=|\psi^{(2)}|^2$.
Then Eq. (\ref{hm_dual}) for $N=2$, $e \neq 0$ has the same form
as the dual gauge field correlator for the case $N=1, e=0$, which exhibits a dual 
Higgs phenomenon \cite{hove2000}. Thus, for $N=2, e \neq 0$,
 $\ve m^{(1)} - \ve m^{(2)}$ behaves  as vortices for $N=1, e=0$,
{\it i.e. it is a superfluid mode arising out of superconducting condensates.}
A nonzero  $m_{\Delta \ve h}$ is {\it produced} by disordering $\theta^{(1)}$ at
$T_{c1}$ while a nonzero  $m_{\ve A}$ is {\it destroyed} by disordering
$\theta^{(2)}$ at $T_{c2}$.

We have  analysed  the $N$-color London model Eq. (\ref{villain}) in vortex representation 
Eqs. (\ref{vortex_action}) and (\ref{potential}). The dual theory is given by Eqs. (\ref{dual2}) 
and (\ref{dual_action}). For $N=2$, we have performed large scale Monte Carlo simulations 
computing \textit{i)} critical exponents $\alpha$ and $\nu$, \textit{ii)} gauge field and dual 
gauge field correlators, \textit{iii)} the corresponding masses, and \textit{iv)} critical 
couplings using FSS. For $\psi^{(1)} \neq \psi^{(2)}$ we find one {\it neutral} low-temperature 
critical point at $T_{\rm c 1}$, and one {\it charged} critical point at 
$T_{\rm c 2} > T_{\rm c1}$. For general $N$, a Higgs mass $m_{\ve A}$ is 
generated at the highest critical temperature, followed by $N-1$ neutral fixed points 
as the temperature is lowered.

These results  apply to electronic and protonic condensates in  liquid metallic 
hydrogen under extreme pressure. Estimates exist for $T_{\rm c2}$ for such systems, 
$T_{\rm c2} \approx 160 {\rm K}$ \cite{Ashcroft1999}, and hence  $T_{\rm c1} \approx
0.1 {\rm K}$. Hence, in addition to the emergence of the  Meissner effect at 
$T_{\rm c2}$ and a corresponding divergence in the magnetic penetration length 
$\lambda \sim |1-T/T_{\rm c2}|^{-\nu/(2-\eta_{\ve A})}$ \cite{Blatter}, there will 
also be a novel effect, namely a low-temperature anomaly in the magnetic penetration 
length $\lambda \sim 1/m_{\ve A}$ at $T_{\rm c1}$, cf. Fig. (\ref{gauge_mass}),
due to the appearance of superfluid modes arising from superconducting condensates.

Work was supported by the Norwegian High Performance
Computing Program, and by the Research Council of Norway, Grant
Nos. 158518/431, and 158547/431 (NANOMAT). We acknowledge 
communications with K. B\o rkje, H. Kleinert, O. Motrunich, 
S. Sachdev, Z. Tesanovic, and A. Vishwanath.

\end{document}